\documentclass[journal, onecolumn, pagenumbers]{IEEEtran}
\usepackage{listings}
\usepackage{amsmath}
\usepackage{amsthm}
\usepackage{tikz}
\usepackage{caption}
\usepackage{array}
\usepackage{mdwmath}
\usepackage{multirow}
\usepackage{mdwtab}
\usepackage{eqparbox}
\usepackage{amsfonts}
\usepackage{tikz}
\usepackage{multirow,bigstrut,threeparttable}
\usepackage{amsthm}
\usepackage{array}
\usepackage{bbm}
\usepackage{subfigure}
\usepackage{epstopdf}
\usepackage{mdwmath}
\usepackage{mdwtab}
\usepackage{eqparbox}
\usepackage{tikz}
\usepackage{latexsym}
\usepackage{cite}
\usepackage{amssymb}
\usepackage{bm}
\usepackage{amssymb}
\usepackage{graphicx}
\usepackage{mathrsfs}
\usepackage{epsfig}
\usepackage{psfrag}
\usepackage{setspace}
\usepackage{hyperref}
\usepackage{algorithm}
\usepackage{algpseudocode}
\usepackage{stfloats}

\newtheorem{remark}{Remark}

\newtheorem{theorem}{Theorem}

\newtheorem{lemma}{Lemma} 
\newtheorem{corollary}{Corollary}

\newtheorem{definition}{Definition}

\def \bE {\mathbb{E}}

\def \spo {\mathsf{Poi}}

\def \var {\mathsf{Var}}
\def \spo {\mathsf{Poi}}
\newcommand\argmin{\mathop{\mbox{{\rm argmin}}}\limits}
\newcommand\argmax{\mathop{\mbox{{\rm argmax}}}\limits}
\begin{document}

\title{Estimating the Fundamental Limits is Easier than Achieving the Fundamental Limits}

\author{Jiantao~Jiao,~\IEEEmembership{Student Member,~IEEE},~Yanjun~Han,~\IEEEmembership{Student Member,~IEEE}, Irena Fischer-Hwang,~\IEEEmembership{Student Member}, and Tsachy~Weissman,~\IEEEmembership{Fellow,~IEEE}
\thanks{Jiantao Jiao, Yanjun Han, Irena Fischer-Hwang, and Tsachy Weissman are with the Department of Electrical Engineering, Stanford University, CA, USA. Email: \{jiantao,yjhan,ihwang,tsachy\}@stanford.edu}. 
}
%
%

\date{\today}

\vspace{-10pt}

\maketitle

\begin{abstract}
We show through case studies that it is easier to estimate the fundamental limits of data processing than to construct explicit algorithms to achieve those limits. Focusing on binary classification, data compression, and prediction under logarithmic loss, we show that in the finite space setting, when it is possible to construct an estimator of the limits with vanishing error with $n$ samples, it may require at least $n\ln n$ samples to construct an explicit algorithm to achieve the limits. 
\end{abstract}


\section{Introduction}

Suppose there exist three machine learning experts that would like to understand the fundamental limits of classification (Bayes error)\cite{Devroye--Gyorfi--Lugosi1996probabilistic} for a specific dataset. Since the true distribution that generates the data is unknown, they take three different approaches:

\begin{enumerate}
\item Expert $A$: given empirical training samples, produce an estimate of the Bayes error that is (near) optimal statistically; 
\item Expert $B$: construct a (near) optimal classifier based on the training sample, and then use its performance on the test set (may have infinite size) to estimate the Bayes error;
\item Expert $C$: use the training error of a (near) optimal classification algorithm to estimate the Bayes error. 
\end{enumerate}

We ask the question: are there any fundamental differences between experts $A,B$, and $C$? Evidently, expert $A$ is not constrained by any specific approaches as experts $B$ and $C$ are, but if $B$ and $C$ are using (near) optimal classification algorithms, would $B$ or $C$ achieve the same performance of $A$ if $A$ chooses to act optimally? 

Similar situations arise in the understanding of fundamental limits of data compression and sequential prediction under logarithmic loss, which is given by the Shannon entropy rate~\cite{Cover--Thomas2006}. In this situation, there could exist four different experts:  
\begin{enumerate}
\item $A$: would like to estimate the limits of compression (near) optimally;
\item $B$: would like to construct a predictor based on training samples and use its prediction accuracy under logarithmic loss on the test set (may have infinite size) to estimate the limits;
\item $C$: would like to use the training error of a (near) optimal sequential predictor to estimate the limits;
\item $D$: would like to construct a (near) optimal data compressor and use its normalized code length to estimate the limits. 
\end{enumerate}

In this situation, are there any fundamental differences between the tasks of these four experts? 

The main message from this paper is that there \emph{exist} significant differences between the difficulties of tasks of these experts in general. In particular, expert $A$'s task is generally significantly easier than that of others. More precisely, when there exist algorithms for expert $A$ to achieve vanishing error with $n$ samples, it may require at least $n\ln n$ samples for other experts to achieve the same performance. 

It may be unexpected that the differences between those approaches could be so significant. Indeed, it has been a long tradition in the information theory and machine learning community to understand the fundamental limits of prediction by iteratively improving existing prediction algorithms and computing the algorithm performance on the test set as benchmarks. However, we argue that even if we have a test size of infinite size, this approach may still be strictly significantly sub-optimal when compared with approaches that directly estimate the fundamental limits without constructing a prediction algorithm explicitly. 

\subsection{Background}

In statistics and machine learning, the \emph{fundamental limits} usually refer to the optimal performance achievable by a certain class of schemes. Various statistical functional are used to quantify the fundamental limits, such as the KL divergence as the Stein exponent~\cite{Lehmann--Romano2005}, the Chernoff information~\cite{Cover--Thomas2006}, the total variation distance~\cite{Lehmann--Romano2005}, and the Shannon entropy~\cite{Cover--Thomas2006}. 

Certain functionals are motivated by asymptotic analysis, such as the Stein exponent and the Chernoff information, while others are exact finite sample fundamental limits, such as the total variation distance and the Shannon entropy. We focus on the \emph{exact} fundamental limits in this paper. It turns out that a variety of the well-known fundamental limits in data processing comes from the \emph{Bayes envelope} computation, which we introduce briefly below. 

Suppose we have a random variable $Z \sim P, Z\in \mathcal{Z}$. For simplicity, we focus on the finite alphabet setting, i.e., the cardinality $|\mathcal{Z}|$ of space $\mathcal{Z}$ satisfies $|\mathcal{Z}|<\infty$. We would like to predict $Z$ using an arbitrary predictor $\hat{Z} \in \hat{\mathcal{Z}}$. Note that it is not necessary that $\mathcal{Z} = \hat{\mathcal{Z}}$. Under loss function $L(Z,\hat{Z})$, we define the \emph{Bayes envelope} (also called \emph{generalized entropy}) as follows:
\begin{align}
U(P_Z) & = \inf_{\hat{Z}} \mathbb{E}_P[L(Z,\hat{Z})] \\
& = \inf_{\hat{Z}} \int L(Z,\hat{Z}) dP_Z.
\end{align}

In other words, the Bayes envelope $U(P_Z)$ quantifies the optimal performance one can ever achieve under loss function $L(Z;\hat{Z})$ if the predictor $\hat{Z}$ is independent of the random variable of interest $Z$. We have suppressed the dependence of $U(P_Z)$ on the loss function $L$. 

The Bayes envelope satisfies the following properties:
\begin{enumerate}
\item It is a concave function of $P_Z$. Indeed, it is defined as the infimum over a family of linear functionals of $P_Z$, which is in general concave~\cite[Chap. 3.2.3]{boyd2004convex}. 
\item Suppose one observes $Z_1,Z_2,\ldots,Z_n \stackrel{\mathrm{i.i.d.}}{\sim} P_Z$, and constructs a predictor $\hat{Z} = \hat{Z}(Z_1,Z_2,\ldots,Z_n)$. Then,
\begin{align} \label{eqn.achieveupperbias}
\mathbb{E}_{P_{Z_1,Z_2,\ldots,Z_n,Z}}\left[ L(Z,\hat{Z}) \right] - U(P_Z) & \geq 0.
\end{align}
This follows from the tower property of conditional expectation and the fact that $Z$ is independent of $\{Z_1,Z_2,\ldots,Z_n\}$. 
\end{enumerate}

These properties reinforces the significance of the Bayes envelope as a measure of fundamental limits. We focus on two specific cases of the Bayes envelope, which corresponds to prediction under logarithmic loss (which is also intimately connected to data compression), and binary classification. 

\begin{enumerate}
\item Prediction under logarithmic loss: the logarithmic loss $L_{\lg}(z, \hat{P})$ is defined as
\begin{align}\label{eqn.loglossdefinition}
L_{\lg}(z,\hat{P}) & = \lg \frac{1}{\hat{P}(x)}, x\in \mathcal{Z}.
\end{align}
Here $\lg$ denotes $\log_2$. In other words, the reconstruction $\hat{P} \in \hat{\mathcal{Z}}$ lies in the space of probability measures $\hat{\mathcal{Z}}$ on $\mathcal{Z}$. The Bayes envelope in this case reduces to 
\begin{align}
U(P_Z) = H(P_Z) = \sum_{z\in \mathcal{Z}} P_Z(z)\lg \frac{1}{P_Z(z)},
\end{align}
where $H(P_Z)$ is the Shannon entropy. It follows from the nonnegativity of the KL divergence. 
\item Binary classification: in binary classification, we have a random vector $Z = (X,Y) \in \mathcal{Z} = \mathcal{S} \times \{0,1\}$, where $X$ represents the feature, $Y$ represents the label. We use the Hamming loss $\mathbbm{1}(t(X)\neq Y)$ to quantify the performance of any classifier $t: \mathcal{S} \mapsto \{0,1\}$. The Bayes envelope in this setting is reduced to~\cite[Chap. 2]{Devroye--Gyorfi--Lugosi1996probabilistic}:
\begin{align}
U(P_Z) & = \mathbb{E}_P\left[ \min\{ \eta(X), 1-\eta(X)\} \right],
\end{align}
where $\eta(x) = P(Y = 1|X = x)$. If we further know that $P(Y =1) = P(Y = 0) = \frac{1}{2}$, then
\begin{align}
U(P_Z) & = \frac{1}{2} - \frac{1}{4} L_1( P_{X|Y = 0}, P_{X|Y = 1} ),
\end{align}
where $L_1(P,Q)$ denotes the $L_1$ distance between two probability measures defined as $L_1(P,Q) = \int |p(x) - q(x)|d\nu$, and $p(x) = \frac{dP}{d\nu}, q(x)= \frac{dQ}{d\nu}$. 
\end{enumerate}

Now we formally define the two distinct problems of estimating fundamental limits and achieving fundamental limits. 
\begin{definition}[Estimating the fundamental limits] \label{def.estimatefundamentallimits}
Given $Z_1,Z_2,\ldots,Z_n \stackrel{\mathrm{i.i.d.}}{\sim} P_Z$, the problem of \emph{estimating fundamental limits} is defined as solving the following minimax problem:
\begin{align}
R_{\mathsf{EST}}(\mathcal{D}, L, n) & =  \min_{\hat{U}(Z_1,Z_2,\ldots,Z_n)} \sup_{P_Z \in \mathcal{D}} \mathbb{E}_P|\hat{U} - U(P_Z)|,
\end{align}
where the supremum is over a collection of probability measures on $\mathcal{Z}$, denoted as $\mathcal{D}$, and the infimum is over all possible estimators of $U(P_Z)$ given $n$ empirical samples. 
\end{definition}

\begin{definition}[Achieving the fundamental limits]\label{def.achievingfundamentallimits}
Given $Z_1,Z_2,\ldots,Z_n\stackrel{\mathrm{i.i.d.}}{\sim} P_Z$, the problem of achieving the fundamental limit is defined as solving the following minimax problem:
\begin{align}
R_{\mathsf{ACH}}(\mathcal{D}, L,n) = \inf_{\hat{Z}(Z_1,Z_2,\ldots,Z_n)} \sup_{P_Z\in \mathcal{D}} \left( \mathbb{E}_{P_{Z_1,Z_2,\ldots,Z_n,Z}}\left[ L(Z,\hat{Z}) \right] - U(P_Z) \right),
\end{align}
where the supremum is over a collection of probability measures on $\mathcal{Z}$, denoted as $\mathcal{D}$, and the infimum is over all possible predictors of $Z$ given $n$ empirical samples. 
\end{definition}

Since we have assumed that $Z$ lies in a finite alphabet, there exists a natural plug-in estimator for the fundamental limit $U(P_Z)$: the estimator $U(P_n)$, where $P_n$ is the empirical distribution of $Z_1,Z_2,\ldots,Z_n$. Interestingly, $U(P_n)$ is always a lower estimate of $U(P_Z)$ on expectation: 
\begin{align}\label{eqn.plugindownbias}
\mathbb{E}_P[U(P_n)] \leq U(P_Z),
\end{align}
which follows from Jensen's inequality. Analogously, we can define the performance of this plug-in approach as follows:
\begin{definition}[Plug-in approach of estimating fundamental limits]\label{def.pluginachievefundamental}
Given $Z_1,Z_2,\ldots,Z_n\stackrel{\mathrm{i.i.d.}}{\sim} P_Z$, denote by $P_n$ the empirical distribution of $Z_1,Z_2,\ldots,Z_n$. In other words, $P_n(z) = \sum_{i =1}^n \frac{\mathbbm{1}(Z_i = z)}{n}$ for any $z\in \mathcal{Z}$. The problem of the plug-in approach $U(P_n)$ in estimating $U(P_Z)$ is defined as the following worst case risk:
\begin{align}
R_{\mathsf{PLU}}(\mathcal{D}, L, n) = \sup_{P_Z\in \mathcal{D}} \mathbb{E}_P | U(P_n) - U(P_Z)|,
\end{align}
where the supremum is over a collection of probability measures on $\mathcal{Z}$, denoted as $\mathcal{D}$. 
\end{definition}

\emph{Notation:} for non-negative sequences $a_\gamma,b_\gamma$, we use the notation $a_\gamma \lesssim  b_\gamma$ to denote that there exists a universal constant $C$ such that $\sup_{\gamma } \frac{a_\gamma}{b_\gamma} \leq C$, and $a_\gamma \gtrsim b_\gamma$ is equivalent to $b_\gamma \lesssim a_\gamma$. Notation $a_\gamma \asymp b_\gamma$ is equivalent to $a_\gamma \lesssim  b_\gamma$ and $b_\gamma \lesssim  a_\gamma$. Notation $a_\gamma \gg b_\gamma$ means that $\liminf_\gamma \frac{a_\gamma}{b_\gamma} = \infty$, and $a_\gamma \ll b_\gamma$ is equivalent to $b_\gamma \gg a_\gamma$. We write $a\wedge b=\min\{a,b\}$ and $a\vee b=\max\{a,b\}$. We use $\ln$ to denote $\log_e$ and $\lg$ to denote $\log_2$. We denote the vector $(X_i,X_{i+1},\ldots,X_k)$ by $X_i^k$. If $i>k$, $X_i^k = \emptyset$. We denote by $D(P\|Q) = \sum_{x\in \mathcal{S}} P(x) \lg \frac{P(x)}{Q(x)}$ the Kullback--Leibler divergence between $P$ and $Q$. 

\subsection{Main results}

\subsubsection{Effective sample size enlargement}

We provide explicit solutions of the three aforementioned problems for a variety of $\mathcal{D}$ and $L$ that include prediction under logarithmic loss and the binary classification case. The main findings may be summarizes by the following statement:
\begin{theorem}[``Informal'']\label{thm.informalmaincontributions}
Estimating the fundamental limits optimally is easier than achieving the fundamental limits, and using the plug-in rule to estimate. Concretely, for a variety of $\mathcal{D},L$, there exists an \emph{effective sample size enlargement phenomenon}:
\begin{align}
R_{\mathsf{EST}}(\mathcal{D},L,n) & \approx R_{\mathsf{ACH}}(\mathcal{D},L, n\ln n) \approx R_{\mathsf{PLU}}(\mathcal{D},L,n\ln n).
\end{align}
\end{theorem}

In other words, the performance of the optimal scheme in estimating the fundamental limits with $n$ samples is essentially that of the optimal scheme in achieving the fundamental limits with $n\ln n$ samples, which is also essentially that of the plug-in approach in estimating the fundamental limits with $n\ln n$ samples. It is an interesting fact that the two distinct problems, i.e., achieving the fundamental limits and estimating the fundamental limits using plug-in approach, enjoy essentially the same performance, while the optimal approach in estimating the limits is far better. The logarithmic sample size enlargement phenomenon between $R_{\mathsf{PLU}}$ and $R_{\mathsf{EST}}$ was identified in~\cite{Jiao--Venkat--Han--Weissman2015minimax} and named \emph{effective sample size enlargement}. Theorem~\ref{thm.informalmaincontributions} provides a generalized view of this phenomenon, which includes $R_{\mathsf{ACH}}$ in the arena. 

\begin{remark}[Bias is the dominating factor]
The problem of achieving the fundamental limits can be cast as approaching the limits from the top: indeed, (\ref{eqn.achieveupperbias}) shows us that the average performance of the predictor is always an upper bound on the Bayes envelope. In contrast, the plug-in approach can be cast as approaching the limits from the bottom: indeed, (\ref{eqn.plugindownbias}) shows that the expectation of the plug-in estimator is always a lower bound on the Bayes envelope. However, the optimal approach in estimating the Bayes envelope does not suffer from any type of bias constraints, which turns out to be the key reason why the optimal approach achieves a logarithmic gain in the performance. It is interesting to see that the bias constraints in the other two problems are so severe that within the constraints one cannot achieve the optimal performance in estimating the fundamental limits. 
\end{remark}

Concretely, we collect various results scattered in the literature that follow the theme of Theorem~\ref{thm.informalmaincontributions} for prediction under logarithmic loss and binary classification, and strengthen the existing results by providing a refined analysis in the binary classification setting. 

The main technical theorem in this paper is the following. 
\begin{theorem}\label{thm.fixedq3values}
Consider the case of 
\begin{enumerate}
\item $Z = (X,Y) \in \mathcal{S} \times \{0,1\}, |\mathcal{S}| = S$;
\item $\mathcal{D} = \mathcal{P}(Q, \frac{1}{2})$;
\item $\hat{Z} = t$, where $t: \mathcal{S} \mapsto \{0,1\}$ is an arbitrary classifier; 
\item $L(Z,\hat{Z}) = \mathbbm{1}(Y \neq t(X))$;
\end{enumerate}
where $\mathcal{P}(Q, \frac{1}{2})$ denotes the space of probability measures on $(X,Y)\in \mathcal{S} \times \{0,1\}$ that satisfies $P(Y = 0) = P(Y = 1) = \frac{1}{2}$, and $P_{X|Y = 1} = Q$, and $Q = (q_1,q_2,\ldots, q_S)$ is a fixed distribution. 

Given $n$ i.i.d. samples $X_1,X_2,\ldots,X_n$ from $P_{X|Y = 0}$, if 
\begin{align}
\ln S \lesssim \ln n \lesssim \ln \left( \sum_{i = 1}^S \sqrt{q_i} \wedge q_i \sqrt{n\ln n} \right),
\end{align}
then
\begin{align}
R_{\mathsf{EST}}(\mathcal{D}, L, n) & \asymp \sum_{i = 1}^S q_i \wedge \sqrt{\frac{q_i}{n\ln n}} \\
R_{\mathsf{ACH}}(\mathcal{D}, L, n) & \asymp \sum_{i = 1}^S q_i \wedge \sqrt{\frac{q_i}{n}} \\
R_{\mathsf{PLU}}(\mathcal{D}, L, n) & \asymp \sum_{i = 1}^S q_i \wedge \sqrt{\frac{q_i}{n}}. 
\end{align}
Moreover, the maximum likelihood classifier 
\begin{align}\label{eqn.mleknownq}
t_{\mathsf{MLE}}(x) = 1 - \mathbbm{1}\left( \frac{\sum_{j = 1}^n \mathbbm{1}(X_j = x)}{n} > q_x \right)
\end{align}
does not achieve the rate of $R_{\mathsf{ACH}}(\mathcal{D}, L, n)$. The classifier defined as 
\begin{align}
t_{Q}(x) = 1 - \mathbbm{1}\left( \left \{  \frac{\sum_{j = 1}^n \mathbbm{1}(X_j = x)}{n} > q_x \right \} \bigcup \left\{ q_x < \frac{1}{n} \right \} \right) 
\end{align}
achieves the rate of $R_{\mathsf{ACH}}(\mathcal{D}, L, n)$.
\end{theorem}

\begin{remark}
The results on $R_{\mathsf{EST}}(\mathcal{D}, L, n)$ and $R_{\mathsf{PLU}}(\mathcal{D}, L, n) $ follow from~\cite{Jiao--Han--Weissman2016l1distance}. The key contribution of this paper is the solution of $R_{\mathsf{ACH}}(\mathcal{D}, L, n)$, whose upper and lower bounds prove to be non-trivial. Indeed, it may seem weird that the maximum likelihood classifier does not achieve the minimax regret in achieving the fundamental limits, while $R_{\mathsf{ACH}}(\mathcal{D}, L, n)$ and $R_{\mathsf{PLU}}(\mathcal{D}, L, n)$ are of the same order. One intuitive explanation might be the following. Given the knowledge of $Q$, if we know $q_x < \frac{1}{n}$, then we should always classify the symbol $x$ into class zero. It is because even if the symbol $x$ indeed comes from class one, the regret caused by this wrong classification is well controlled since $q_x$ itself is very small. 
\end{remark}

The following corollary is immediate by taking $Q$ to be the uniform distribution on $\mathcal{S}$. 
\begin{corollary}
It requires $n\gg \frac{S}{\ln S}$ samples to achieve vanishing $R_{\mathsf{EST}}(\mathcal{D}, L, n) $ in the worst case, while it requires $n\gg S$ samples to achieve vanishing $R_{\mathsf{ACH}}(\mathcal{D}, L, n) $ or $R_{\mathsf{PLU}}(\mathcal{D}, L, n) $ in the worst case.
\end{corollary}

\begin{remark}
It is aesthetically pleasing to see that the \emph{effective sample size enlargement} phenomenon holds precisely for every $Q$, as shown in Theorem~\ref{thm.fixedq3values}.
\end{remark}

\begin{figure}[h]
\centering
\begin{minipage}{.48\textwidth}
  \centering
  \includegraphics[width=.8\linewidth]{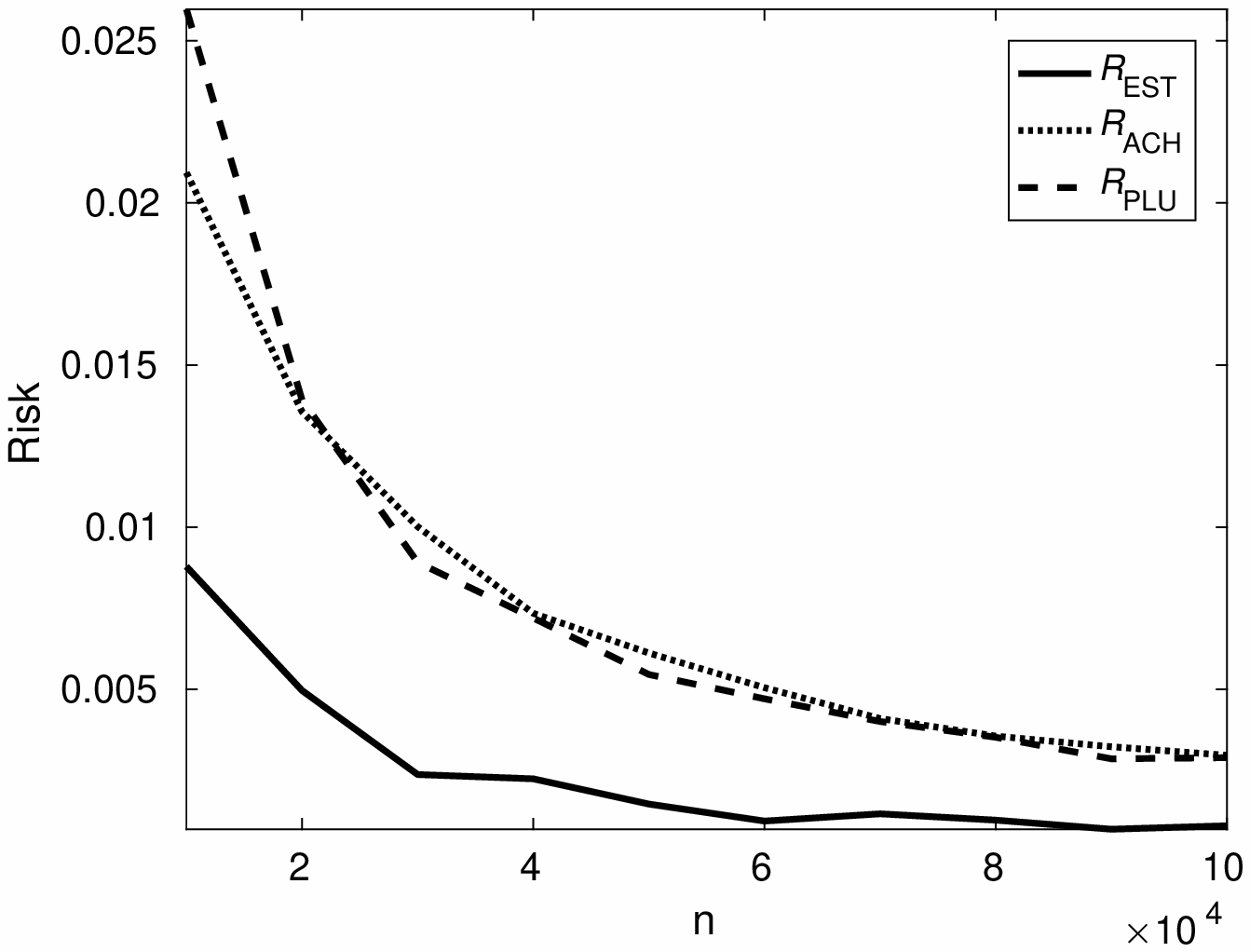}
  \captionof{figure}{Comparing the estimate ($R_{\mathsf{EST}}$) with the plug-in approach $R_{\mathsf{PLU}}$ and the achiever $R_{\mathsf{ACH}}$ of the fundamental limit for an unknown $P$ and known $Q$.}
  \label{fig:test1}
\end{minipage}\hfill%
\begin{minipage}{.48\textwidth}
  \centering
  \includegraphics[width=.8\linewidth]{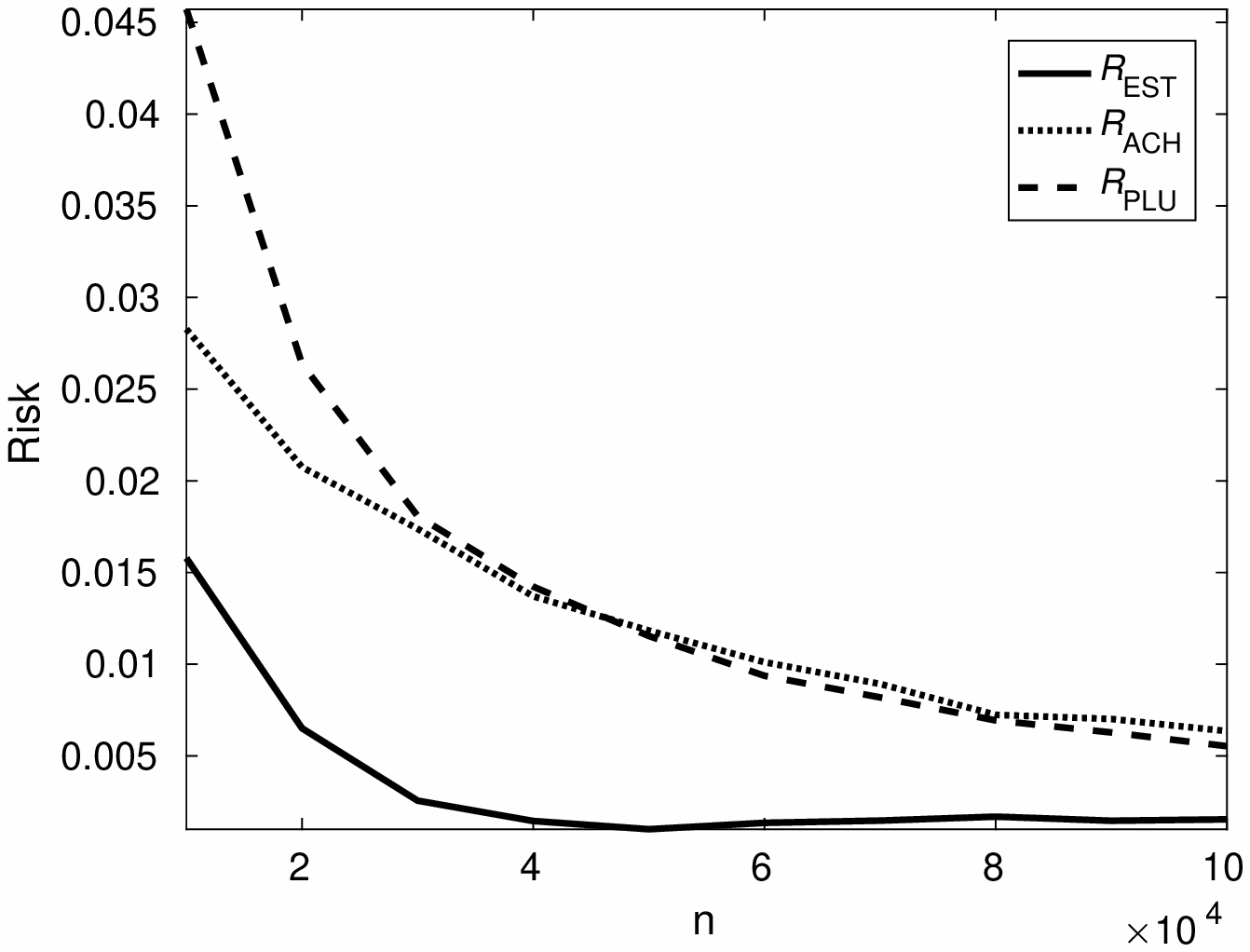}
  \captionof{figure}{Comparing the estimate ($R_{\mathsf{EST}}$) with the plug-in approach $R_{\mathsf{PLU}}$ and the achiever $R_{\mathsf{ACH}}$ of the fundamental limit for both $P$ and $Q$ unknown.}
  \label{fig:test2}
\end{minipage}
\end{figure}

Fig.~\ref{fig:test1} compares the optimal ($R_{\mathsf{EST}}$) and plug-in approaches ($R_{\mathsf{PLU}}$) of estimating fundamental limits with the fundamental limit achiever ($R_{\mathsf{ACH}}$) for the case of $P$ an unknown Zipf distribution with distribution parameter $\beta=0.3$, and $Q$ a known uniform distribution over support size $S=1,000$. This case was simulated for sample sizes $n = 10,000:100,000$ at increments of 10,000 samples, with each simulation case repeated for 20 iterations. Fig.~\ref{fig:test2} compares the estimation approaches with the fundamental limit achiever for the case of both $P$ and $Q$ unknown distributions over a support size of $S=1,000$. As with the simulations shown in Fig.~\ref{fig:test1}, $P$ is a Zipf distribution with distribution parameter $\beta=0.3$, while $Q$ is a uniform distribution. Simulations were performed for sample sizes $n = 10,000:100,000$ at increments of 10,000 samples, with each simulation case repeated for 20 iterations.  

\subsubsection{Connections between data compression and entropy estimation}

A (source) code $C_n: \mathcal{S}^n \mapsto \{0,1\}^*$ is defined as an injective mapping from the set $\mathcal{S}^n$ of all sequences of length $n$ over the finite alphabet $\mathcal{S}$ of size $S = |\mathcal{S}|$ to the set $\{0,1\}^*$ of all binary sequences. We consider here only fixed-to-variable uniquely decodable coding. For a given code $C_n$, we let $L(C_n, x_1^n)$ be the code length for $x_1^n$. For any code $C_n$, we know~\cite[Thm. 5.3.1]{Cover--Thomas2006} that
\begin{align}
\mathbb{E}\left[ L(C_n, X_1^n) \right] \geq  H_n(P),
\end{align}
where 
\begin{align}
H_n(P) = \sum_{x_1^n} P(x_1^n) \lg \frac{1}{P(x_1^n)}
\end{align}
is the Shannon entropy of the random variables $X_1^n$. We consider codes that are designed for a certain family of sources $\mathcal{D}$ that generates real data. 

We introduce the normalized average minimax redundancy $\bar{R}_n(\mathcal{D})$ as
\begin{align}
\bar{R}_n(\mathcal{D}) \triangleq \min_{C_n \in \mathcal{C}} \sup_{P \in \mathcal{D}} \frac{1}{n} \left( \mathbb{E}_P [L(C_n,X_1^n) -H_n(P) \right),
\end{align}
where $\mathcal{C}$ denotes the set of all uniquely decodable fixed-to-variable codes. To simplify the argument we adopt the convention to use the continuous approximation of the code length $L(C_n,x_1^n)$, which is $\lg \frac{1}{Q(x_1^n)}$. We introduce the corresponding continuous redundancy as follows:
\begin{align}\label{eqn.minimaxredundancy}
\tilde{R}_n(\mathcal{D}) \triangleq \inf_Q \sup_{P \in \mathcal{D}} \frac{1}{n} D(P_{X^n} \| Q_{X^n}). 
\end{align}

The following result is immediate. \footnote{Indeed, for any code $C_n$, we can define $Q(x^n) = \frac{2^{-L(C_n,x_1^n)}}{\sum_{x_1^n} 2^{-L(C_n, x_1^n)}}$, which leads to
\begin{align}
D(P_{X^n} \| Q_{X^n}) & = \lg\left( \sum_{x_1^n} 2^{-L(C_n, x_1^n)} \right) + \mathbb{E}_P L(C_n, X_1^n) - H_n(P) \\
& \leq \mathbb{E}_P L(C_n, X_1^n) - H_n(P),
\end{align}
where we used Kraft's inequality which states that $\sum_{x_1^n} 2^{-L(C_n,x_1^n)}\leq 1$~\cite[Thm. 5.2.1]{Cover--Thomas2006}. On the other hand, for the optimal distribution $Q$, we use the Shannon code $C_n$ which satisfies $L(C_n, x_1^n) = \lceil \lg \frac{1}{Q(x_1^n)} \rceil$ and achieves
\begin{align}
\mathbb{E}_P L(C_n, X_1^n) - H_n(P) & \leq \mathbb{E}_P \left[ \frac{1}{\lg Q(X_1^n)} + 1\right] - H_n(P) \\
& \leq D(P_{X^n} \| Q_{X^n}) + 1. 
\end{align}}
\begin{align}
\tilde{R}_n(\mathcal{D})\leq \bar{R}_n(\mathcal{D}) \leq \tilde{R}_n(\mathcal{D}) + \frac{1}{n}.  
\end{align}

Consider the case of $\mathcal{D} = \mathcal{D}_0(S)$, where $\mathcal{D}_0(S)$ denotes the space of probability measures on $\mathcal{S}$ with alphabet size $S$. The minimax redundancy for memoryless sources $\tilde{R}_n(\mathcal{D}_0(S))$ has been studied extensively in the literature~\cite{Rissanen1986stochastic, xie1997minimax, xie2000asymptotic, drmota2004precise}. 

We have the following result whose upper bound follows from~\cite{Szpankowski--Weinberger2012minimax}. 
 
\begin{theorem}\label{thm.iidcompression}
Suppose $S = \alpha n$, where $\alpha\in \left(0, \frac{e}{2\pi} \right)$ is a constant. Then, the minimax redundancy
\begin{align}
\liminf_n \tilde{R}_n(\mathcal{D}_0(S)) & \geq \frac{\alpha}{2} \lg \left( \frac{e}{2\pi \alpha} \right) \\
\limsup_n \tilde{R}_n(\mathcal{D}_0(S)) & \leq \lg B_\alpha,
\end{align}
where $B_\alpha = \alpha C_\alpha^{\alpha+2} e^{-\frac{1}{C_\alpha}}, C_\alpha = \frac{1}{2} + \frac{1}{2} \sqrt{1 + \frac{4}{\alpha}}$ were introduced in~\cite{Szpankowski--Weinberger2012minimax}. 
\end{theorem}

\begin{remark}\label{remark.achievehard}
Theorem~\ref{thm.iidcompression} shows that if one would like to use the code length $L(C_n, X_1^n)$ to estimate the corresponding Shannon entropy, it would take at least $n\gg S$ samples, while it only requires $n\gg \frac{S}{\ln S}$ samples to estimate the Shannon entropy for memoryless sources~(Lemma~\ref{lemma.achievelogloss} in Section~\ref{sec.existingresults}). 
\end{remark}

One main reason why we discuss the notion of data compression is that unlike the problem of achieving the fundamental limits, a data compressor leads to an entropy estimator, but a predictor as in Definition~\ref{def.achievingfundamentallimits} does not. It has been a long tradition in various communities to use compression theoretic approaches to estimate the entropy rate, but Remark~\ref{remark.achievehard} suggests that even the best compressor cannot achieve the optimal performance in entropy estimation. In order to illustrate this seemingly weird phenomenon, we now investigate the idea of compression based entropy estimation. 

The idea of using code length to estimate the Shannon entropy can be viewed as using the following estimator 
\begin{align}
\hat{H}_{\mathrm{Compression}} & = \frac{1}{n} \sum_{i = 1}^n \lg \frac{1}{Q_{X_i|X^{i-1}}(X_i|X^{i-1})},
\end{align}
where $Q$ is a specific coding distribution on the space $\mathcal{S}^n$ used by the data compressor. The quantity $Q_{X_i|X^{i-1}}(X_i|X^{i-1})$ is a random variable where the randomness is induced by the random variables $X_1,X_2,\ldots, X_i$ which are i.i.d. with distribution $P$. The data compression type estimate is always biased upwards in the sense that 
\begin{align}\label{eqn.compressionupper}
\mathbb{E}[\hat{H}_{\mathrm{Compression}}] - \frac{1}{n} \mathbb{E}_P \lg \frac{1}{P_{X^n}(X^n)} & = \frac{1}{n}D(P_{X^n} \| Q_{X^n}) \geq 0.  
\end{align}

The plug-in approach~(Def.~\ref{def.pluginachievefundamental}) in entropy estimation can also be viewed in the similar form:
\begin{align}
H(P_n) & = \frac{1}{n} \sum_{i = 1}^n \lg \frac{1}{P_n(X_i)} \\
& = \min_{P \in \mathcal{D}_0(S)}  \frac{1}{n} \sum_{i = 1}^n \lg \frac{1}{P(X_i)}, 
\end{align}
where $P_n$ is the empirical distribution, and in the last step we used the fact that the empirical distribution maximizes the likelihood~\cite[Chap. 2.1]{owen2001empirical}. It is clear that 
\begin{align}
\mathbb{E}[\hat{H}_{\mathrm{Compression}}] & \geq H(P) \geq \mathbb{E}[H(P_n)]. 
\end{align}

The key observation is that $n$-fold product distribution $P_n\otimes P_n \otimes \ldots \otimes P_n$ \emph{cannot} be used as coding distribution since it is dependent on the empirical data and thus unknown to the decoder. The constraint on data compression forces one to use a distribution $Q$ that is independent of the empirical data. However, if we only care about estimating the fundamental limits, we can in fact use 
\begin{align}\label{eqn.arbitrayg}
\frac{1}{n} \sum_{i = 1}^n \hat{g}(X_i),
\end{align}
where $\hat{g}$ is any random function depending on the empirical data $X_1^n$. It turns out that among estimators of type~(\ref{eqn.arbitrayg}) there exist minimax rate-optimal estimators that achieve $R_{\mathsf{EST}}(\mathcal{D}_0(S),L,n)$ up to universal constants~\cite{Valiant--Valiant2011,Valiant--Valiant2013estimating,Jiao--Venkat--Han--Weissman2015minimax,Wu--Yang2014minimax}. It is also interesting that not only plugging-in the empirical distribution fails to achieve the minimax rates in estimating entropy, plugging-in the Dirichlet prior smoothed distribution estimates~\cite{Han--Jiao--Weissman2015adaptive} also fails.

The rest of the paper is organized as follows. The collection of existing results in the literature on binary classification and prediction under logarithmic loss is collected in Section~\ref{sec.existingresults}. The auxiliary lemmas used in the proofs of the main results in gathered in Section~\ref{sec.auxiliarylemmas}. We prove the main theorems in Section~\ref{sec.proofsofmaintheorems}. The proofs of the auxiliary lemmas are presented in Section~\ref{sec.proofsofauxiliarylemmas}.


%

\section{Existing results on classification and prediction under logarithmic loss}\label{sec.existingresults}

\subsection{Classification}

Specialize the general definition of achieving the fundamental limits~(Def.~\ref{def.achievingfundamentallimits}) to the case of
\begin{enumerate}
\item $Z = (X,Y) \in \mathcal{S} \times \{0,1\}, |\mathcal{S}| = S$;
\item $\mathcal{D} = \text{all probability measures on }\mathcal{S}\times \{0,1\}$;
\item $\hat{Z} = t$, where $t: \mathcal{S} \mapsto \{0,1\}$ is an arbitrary classifier; 
\item $L(Z,\hat{Z}) = \mathbbm{1}(Y \neq t(X))$.
\end{enumerate}
It is clear that the problem of solving $R_{\mathsf{ACH}}(\mathcal{D},L,n)$ is nothing but the standard minimax regret problem of binary classification in statistical learning theory~\cite{massart2006risk}. Indeed, 
\begin{align}
R_{\mathsf{ACH}}(\mathcal{D},L,n) = \inf_{t} \sup_{P \in \mathcal{D}} \mathbb{E}[\ell(t^*,t)], 
\end{align} 
where
\begin{align}
\ell(t^*,t) & = P(Y \neq t(X)) - P(Y\neq t^*(X)),
\end{align}
and $t^*$ is the Bayes classifier defined as $t^*(x) = \mathbbm{1}(\eta(x)\geq 1/2)$, where $\eta(x) = P(Y = 1|X = x)$.

Let $\mathcal{F} = \{\mathbbm{1}(A): A\in 2^{\mathcal{S}}\}$ be the collection of all possible classifiers, where $2^{\mathcal{S}}$ is the power set of the feature space $\mathcal{S}$ with the size of $2^S$. It is clear that the Bayes classifier $t^*$ belongs to $\mathcal{F}$ and the collection of sets $2^\mathcal{S}$ has the Vapnik--Chervonenkis dimension $S$. Consider the empirical risk minimization (ERM) classifier $\hat{t}$ that is defined as the classifier that minimizes the empirical risk:
\begin{align}
\hat{t} = \argmin_{t\in \mathcal{F}}  \frac{1}{n} \sum_{i = 1}^n \mathbbm{1}(Y_i \neq t(X_i)). 
\end{align}
The ERM classifier is particularly easy to describe in the discrete feature space setting: for every $x\in \mathcal{S}$, we have $\hat{t}(x) = \argmax_j \sum_{i = 1}^n \mathbbm{1}(Y_i = j, X_i = x)$, where $j \in \{0,1\}$.

The following result is well known. 
\begin{lemma}\cite{devroye1995lower}\cite[Sec A4.5]{Vapnik1998statistical}\cite{lugosi2002pattern}\label{lemma.unknownminimaxregret}
Suppose $2\leq S\leq n$. Then, 
\begin{align}
R_{\mathsf{ACH}}(\mathcal{D},L,n) \asymp \sqrt{\frac{S}{n}},
\end{align}
Furthermore, the minimax regret is achieved by the ERM classifier up to a constant. 
\end{lemma}

Lemma~\ref{lemma.unknownminimaxregret} implies that $n_{\mathsf{ACH}}^*(\epsilon;\mathcal{D}, L) \asymp \frac{S}{\epsilon^2}$ for any constant $\epsilon>0$ that is small enough. Since $n_{\mathsf{ACH}}^*(\epsilon;\mathcal{D}, L)$ is proportional to $S$, it suggests intuitively that any classifier would not be able to achieve vanishing worst case regret if it has not seem all the elements in the feature space $\mathcal{S}$ at least once.

\subsection{Prediction under the logarithmic loss}\label{sec.predictionunderlogloss}
Specializing the general definition of achieving the fundamental limits~(Def.~\ref{def.achievingfundamentallimits}) to the case of
\begin{enumerate}
\item $Z = X\in \mathcal{S}, |\mathcal{S}| = S$;
\item $\mathcal{D} = \mathcal{D}_0(S)$;
\item $\hat{Z} \in \mathcal{D}_0(S)$;
\item $L(Z,\hat{Z}) = \lg \frac{1}{\hat{Z}(Z)}$. 
\end{enumerate}
Here the loss function is the logarithmic loss defined in~(\ref{eqn.loglossdefinition}), and $\mathcal{D}_0(S)$ denotes the space of probability measures on $\mathcal{S}$ with alphabet size $S$. 

It is clear that the problem of solving $R_{\mathsf{ACH}}(\mathcal{D},L,n)$ is nothing but estimating the distribution $P_X$ under the KL divergence loss. Indeed, we have
\begin{align}
R_{\mathsf{ACH}}(\mathcal{D},L,n) = \inf_{\hat{P}(X_1,X_2,\ldots,X_n)} \sup_{P \in \mathcal{D}} \mathbb{E}_P \left[  D( P \| \hat{P}) \right],
\end{align}
where $D(P\|Q) = \sum_{x\in \mathcal{S}} P(x) \lg \frac{P(x)}{Q(x)}$ is the Kullback--Leibler divergence between $P$ and $Q$.  

The following result is well known. 
\begin{lemma}\cite{Paninski2004variational}\cite{Braess--Sauer2004}\cite{Wu--Yang2014minimax}\cite{Jiao--Venkat--Han--Weissman2015minimax} \label{lemma.achievelogloss}
Under the conditions in Section~\ref{sec.predictionunderlogloss},
\begin{enumerate}
\item 
\begin{align}
R_{\mathsf{ACH}}(\mathcal{D}_0(S),L,n) & \begin{cases} = \frac{S-1}{2n} \lg(e) (1+o(1)) & n \gg S \\ \in (0,\infty) & \lim \frac{S}{n} = c, c\in (0,\infty) \\ = (1+o(1))\lg\left(\frac{S}{n} \right) & n \ll S \end{cases};
\end{align}
\item if $n \gtrsim \frac{S}{\ln S}$, 
\begin{align}
R_{\mathsf{EST}}(\mathcal{D}_0(S),L,n) & \asymp \frac{S}{n\ln n} + \frac{\lg S}{\sqrt{n}} 
\end{align}
\item if $n \gtrsim S$,
\begin{align}
R_{\mathsf{PLU}}(\mathcal{D}_0(S),L,n) & \asymp \frac{S}{n} + \frac{\lg S}{\sqrt{n}} . 
\end{align}
\end{enumerate}
\end{lemma}

The following corollary is immediate. 
\begin{corollary}\label{cor.entropyestimation}
It takes $n\gg\frac{S}{\ln S}$ samples to achieve vanishing $R_{\mathsf{EST}}(\mathcal{D}_0(S),L,n)$, while it takes $n\gg S$ samples to achieve vanishing $R_{\mathsf{ACH}}(\mathcal{D}_0(S),L,n)$ and $R_{\mathsf{PLU}}(\mathcal{D}_0(S),L,n)$. 
\end{corollary}

\begin{remark}
It was shown in~\cite{Han--Jiao--Weissman2015adaptive} that the effective sample size enlargement phenomenon between $R_{\mathsf{EST}}(\mathcal{D}_0(S),L,n)$ and $R_{\mathsf{PLU}}(\mathcal{D}_0(S),L,n)$ also holds for more refined subclasses of $\mathcal{D}(H) = \{P \in \mathcal{D}_0(S): H(P)\leq H\}$. 
\end{remark}

\section{Acknowledgment}

We are grateful to Narayana Prasad Santhanam and Wojciech Szpankowski for very helpful discussions about minimax regret in data compression in the large alphabet regime.

\appendices

\section{Auxiliary Lemmas}\label{sec.auxiliarylemmas}

The following lemma presents the Hoeffding bound.
\begin{lemma}\label{lem_hoeffding}
  \cite{Hoeffding1963probability} 
Let $X_1,X_2,\ldots,X_n$ be independent random variables such that $X_i$ takes its value in $[a_i,b_i]$ almost surely for all $i\leq n$. Let $S_n=\sum_{i=1}^n X_i$, we have for any $t>0$,
  \begin{align}
    P\left\{|S_n-\bE[S_n]|\ge t\right\} \le 2\exp\left(-\frac{2t^2}{\sum_{i = 1}^n (b_i-a_i)^2}\right).
  \end{align}
\end{lemma}

 The following lemma gives well-known tail bounds for Poisson and Binomial random variables.
 \begin{lemma}\cite[Exercise 4.7]{mitzenmacher2005probability}\label{lemma.poissontail}
 	If $X\sim \spo(\lambda)$ or $X\sim \mathsf{B}(n,\frac{\lambda}{n})$, then for any $\delta>0$, we have
 	\begin{align}
 	P(X \geq (1+\delta) \lambda) & \leq \left( \frac{e^\delta}{(1+\delta)^{1+\delta}} \right)^\lambda \le e^{-\delta^2\lambda/3}\vee e^{-\delta\lambda/3} \\
 	P(X \leq (1-\delta)\lambda) & \leq  \left( \frac{e^{-\delta}}{(1-\delta)^{1-\delta}} \right)^\lambda \leq e^{-\delta^2 \lambda/2}.
 	\end{align}
 \end{lemma}
 
 \begin{lemma}\cite{Han--Jiao--Weissman2016minimaxdivergence} \label{lem.varprod}
 	For independent random variables $X,Y$ with finite second moment, we have
 	\begin{align}
    \var(XY) = (\bE Y)^2\var(X) + (\bE X)^2 \var(Y) + \var(X)\var(Y).
 	\end{align}
 \end{lemma} 
 
\begin{lemma}\label{lemma.poissoninversebound}\cite[Lemma 34]{Jiao--Han--Weissman2016l1distance}
Suppose $X\sim \spo(\lambda)$ or $X\sim \spo(n, \frac{\lambda}{n})$, where $0< \lambda\leq n$. Then, there exists a universal constant $C>0$ such that
\begin{align}
\mathbb{E}\left[ \frac{1}{X\vee 1} \right] & \leq \frac{C}{\lambda}. 
\end{align}
\end{lemma}

\begin{lemma}\label{lemma.poissondifference}
Suppose $X\sim \spo(\lambda_1)$ or $X\sim \mathsf{B}(n, \frac{\lambda_1}{n})$, where $0\leq \lambda_1 \leq n$. Then, 
\begin{enumerate}
\item if $0\leq \lambda_2 \leq n, \lambda_2 \geq \lambda_1$, we have
\begin{align}
(\lambda_2 - \lambda_1) P(X \geq \lambda_2) & \leq M_1 \lambda_2 \wedge \sqrt{\lambda_2}. 
\end{align}
\item if $\lambda_1\geq  \lambda_2 \geq 1>0$, 
\begin{align}
(\lambda_1 - \lambda_2) P(X \leq \lambda_2) & \leq M_2 \sqrt{\lambda_2}
\end{align} 
Here $M_1>0,M_2>0$ are universal constants that do not depend on $\lambda_1, \lambda_2$, or $n$. 
\end{enumerate}
\end{lemma}

\section{Proofs of main theorems}\label{sec.proofsofmaintheorems}

\subsection{Proof of Theorem~\ref{thm.fixedq3values}}

To simplify notation, we denote
\begin{align}
R & \triangleq P_{X|Y = 0} = (r_1,r_2,\ldots,r_S) \\
Q & \triangleq P_{X|Y = 1} = (q_1,q_2,\ldots,q_S).
\end{align}

Any classifier $t$ is of the form $t(x) = \mathbbm{1}(x \notin \hat{A})$, where $\hat{A}$ is the decision regime for class $0$ of the classifier $t$. We have
\begin{align}
\ell(t^*,t) & = \frac{1}{2}\left(R((\hat{A})^c) + Q(\hat{A}) \right) - P(Y \neq t^*(X)) \\
& = \frac{1}{2}\left(R((\hat{A})^c) + Q(\hat{A}) \right) - \frac{1}{2} + \frac{1}{4} L_1(R,Q) \\
& = \frac{1}{2} \left(  Q(\hat{A}) - R(\hat{A}) + \frac{1}{2} L_1(R,Q) \right) \\
& = \frac{1}{2} \left(  Q(\hat{A}) - R(\hat{A}) + R(A) - Q(A) \right) \\
& =  \frac{1}{2} \left( R(A) - Q(A) - ( R(\hat{A})-Q(\hat{A}) ) \right) \\
& = \frac{1}{2} \sum_{i = 1}^S \left( ( r_i-q_i) \mathbbm{1}(i\in A) - (r_i - q_i) \mathbbm{1}(i \in \hat{A}) \right) \\
& = \frac{1}{2} \sum_{i = 1}^S  ( r_i-q_i) ( \mathbbm{1}(i\in A)-\mathbbm{1}(i \in \hat{A}) ) ,
\end{align}
where the set $A = \{i: r_i > q_i\}$ and we have used the Scheff\'e lemma~\cite[Thm. 5.1]{Devore--Lorentz1993} that $\frac{1}{2} L_1(R,Q) = R(A) - Q(A)$. 

It is clear that the regret $\ell(t^*, t)$ for any $t$ can be written as a function of $R,Q$, and $\hat{A}$. To simply notation we also denote
\begin{align}
\ell(t^*, t) & = \ell(\hat{A}; R,Q) \\
& = \frac{1}{4}L_1(R,Q) - \frac{1}{2}(R(\hat{A}) - Q(\hat{A})),
\end{align}
where the set $A = \{i: r_i>q_i\}$.

The next lemma relates the minimax regret of classification under the Poissonized model of approximate probability distributions to that under the multinomial model of a true probability distribution, where the set of approximate probability distributions is defined by
\begin{align}
\mathcal{D}_0(S,\epsilon) \triangleq \left\{ P = (p_1,p_2,\ldots,p_S): p_i\geq 0, \left| \sum_{i = 1}^S p_i -1 \right| <\epsilon \right \}.
\end{align}

We write the expectation $\mathbb{E}_{R}$ to emphasize that the expectation is taken with respect to the distribution $P$ such that $P_{X|Y = 0} = R, P_{X|Y = 1} = Q, P(Y = 1) = \frac{1}{2}$.

Note that the minimax regret for classification under the multinomial model with known $Q$, $n$ observations on support size $S$ is given by
\begin{align}
R(S,n,Q) = \inf_{\hat{A}} \sup_{R \in \mathcal{D}_0(S,0)} \mathbb{E}_R [\ell(\hat{A};R,Q)],
\end{align}
where the set $\hat{A} = \hat{A}( n\cdot(\hat{r}_1,\hat{r}_2,\ldots,\hat{r}_S),Q)$, where $(\hat{r}_1,\hat{r}_2,\ldots,\hat{r}_S)$ is the empirical distribution of $R$, and the random vector $(n\hat{r}_1,n\hat{r}_2,\ldots,n\hat{r}_S)$ follows multinomial distribution with parameter $n,R$.

Analogously, we can define the corresponding minimax regret for classification under the Poissonized model with known $Q$, support size $S$ and sample size $n$:
\begin{align}
R_P(S,n,Q,\epsilon) & = \inf_{\hat{A}} \sup_{P \in \mathcal{D}_0(S,\epsilon)}\mathbb{E}_R [\ell(\hat{A};R,Q)],
\end{align}
where the set $\hat{A} = \hat{A}( n\cdot(\hat{r}_1,\hat{r}_2,\ldots,\hat{r}_S),Q)$, where $(\hat{r}_1,\hat{r}_2,\ldots,\hat{r}_S)$ is the empirical distribution of $R$, and the random variables $(n\hat{r}_1,n\hat{r}_2,\ldots,n\hat{r}_S)$ are mutually independent with marginal distribution $n\cdot \hat{r}_i \sim \spo(nr_i)$. 

Furthermore, we introduce the Bayes regret with respect to a prior $\mu$ in the Poisson model as 
\begin{align}
R_B(S,n,Q,\mu) & = \inf_{\hat{A}} \int \mathbb{E}_{R_\tau} [\ell(\hat{A}; R_\tau, Q)]\mu(dR).  
\end{align}

We have the following lemmas relating $R_B(S,n,Q,\mu), R(S,n,Q), R_P(S,n,Q,\epsilon)$. 

\begin{lemma}\label{lemma.poissonmultinomial}
For any $S, n\in \mathbb{N}_+,0<\epsilon<1$, we have
\begin{align}
R(S,n(1-\epsilon)/2, Q) & \geq R_P(S,n,Q,\epsilon) - e^{-n(1-\epsilon)/8} - \frac{3\epsilon}{4}. 
\end{align}
\end{lemma}

\begin{lemma}\label{lemma.bayesregret}
If there exists a constant $C>1$ such that 
\begin{align}
\mu \left \{ P: \sum_{i = 1}^S p_i \leq C \right\} = 1,
\end{align}
then
\begin{align}
R_P(S,n,Q,\epsilon) & \geq R_B(S,n,Q,\mu) - C \mu((\mathcal{D}_0(S,\epsilon))^c).
\end{align}
\end{lemma}

We now begin the proof of Theorem~\ref{thm.fixedq3values}. 
\begin{enumerate}
\item Upper bound: 

Given $n$ i.i.d. samples from $R = P_{X|Y = 0}$, we have the empirical distribution $R_n = (\hat{r}_1,\hat{r}_2,\ldots,\hat{r}_S)$. We construct the classifier $t$ whose decision regime $\hat{A} = \{i: \hat{r}_i > q_i\} \cup \{i: q_i < \frac{1}{n}\}$. The regret can be written as:
\begin{align}
\mathbb{E}[ \ell(t^*,t)] & = \frac{1}{2}\left( \sum_{i: r_i>q_i} (r_i - q_i) P(i\notin \hat{A})+ \sum_{i: r_i \leq q_i} (q_i - r_i) P(i\in \hat{A}) \right) \\
& = \frac{1}{2}\left( \sum_{i: r_i>q_i \geq \frac{1}{n}} (r_i - q_i) P( \hat{r}_i \leq q_i) + \sum_{i: r_i \leq q_i} (q_i - r_i) P(\{\hat{r}_i > q_i\} \cup \{q_i < \frac{1}{n}\} ) \right) \\
& \lesssim \frac{1}{2}\left( \sum_{i: r_i>q_i \geq \frac{1}{n}} \frac{1}{n} \sqrt{nq_i}  + \sum_{i: r_i \leq q_i} q_i \wedge \sqrt{\frac{q_i}{n}} \right) \\
& \lesssim \sum_{i = 1}^S q_i \wedge \sqrt{\frac{q_i}{n}},
\end{align}
where in the first inequality step we used Lemma~\ref{lemma.poissondifference}. It is clear that the maximum likelihood approach does not achieve the minimax regret. Indeed, the maximum likelihood approach generates the decision regime $\hat{A}_{\mathsf{MLE}} = \{i: \hat{r}_i >q_i\}$, and for the special case $Q = (0,0,\ldots,0,1)$, it achieves regret exactly
\begin{align}
\frac{1}{2} \left( \sum_{i: q_i = 0} r_i P(\hat{r}_i \leq 0) \right) & = \frac{1}{2} \sum_{i = 1}^{S-1} r_i (1-r_i)^n. 
\end{align}
Taking $r_i = \frac{1}{n}, 1\leq i\leq n$ where $n\geq S-1$, the regret is of order $\frac{S}{n}$, which in general cannot be upper bounded by $\sum_{i = 1}^S q_i \wedge \sqrt{\frac{q_i}{n}} = \frac{1}{\sqrt{n}}$ within a universal constant. 
%
\item Lower bound: 

We first prove the lower bound under the Poisson sampling model, then we use Lemma~\ref{lemma.poissonmultinomial} and Lemma~\ref{lemma.bayesregret} to convert the result back to the multinomial setting. Under the Poisson sampling model, for any $R = P_{X|Y = 0}$, the empirical counts $R_n = (\hat{r}_1,\hat{r}_2,\ldots,\hat{r}_S)$ satisfies $n\cdot \hat{r}_i \sim \spo(nr_i)$ and the random variables $\{r_i:1\leq i\leq S\}$ are mutually independent.

%
%
%

We construct $2^S$ nonnegative vectors indexed by $\tau \in \{-1,1\}^S$, and for each $\tau$, the $i$-th entry of $R_\tau$ is given by
\begin{align}\label{eqn.ptaudefine}
R_\tau(i) = \begin{cases}  q_i + \tau_i c q_i & q_i\leq \frac{1}{n} \\ q_i + \tau_i c \sqrt{\frac{q_i}{n}} & q_i > \frac{1}{n} \end{cases}
\end{align}
for any $1\leq i\leq S$. Here $0<c<1$ is a constant that will be chosen later. Note that $R_\tau$ in general is a nonnegative vector but not a probability distribution.


For any given $\tau = (\tau_1,\tau_2,\ldots,\tau_S)$, let $\tau^j$ denote the $S$-tuple that differs from $\tau$ only on the $j$-th coordinate. We assign the uniform distribution on $\tau$ and denote the induced distribution on $R$ as $\mu$. Note that $\ell(\hat{A}; R,Q) = \frac{1}{4}L_1(R,Q) - \frac{1}{2}(R(\hat{A}) - Q(\hat{A}))$, and $L_1(R_\tau, Q) = c \sum_{i = 1}^S q_i \wedge \sqrt{\frac{q_i}{n}}$. 

We write the expectation $\mathbb{E}_{R_\tau}$ to emphasize that the expectation is taken with respect to the distribution $P$ such that $P_{X|Y = 0} = R_\tau, P_{X|Y = 1} = Q, P(Y = 1) = \frac{1}{2}$. 

We have the Bayes regret
\begin{align}
& \inf_{\hat{A}} \int \mathbb{E}_{R_\tau} [\ell(\hat{A}; R_\tau, Q)]\mu(dR)\\
& \quad = \inf_{\hat{A}} \sum_{\tau} 2^{-S} \mathbb{E}_{R_\tau} \ell(\hat{A};R_\tau,Q) \\
& \quad \geq \frac{c}{4} \sum_{i = 1}^S q_i \wedge \sqrt{\frac{q_i}{n}} - \frac{1}{2} \sup_{\hat{A}} \sum_{\tau} 2^{-S} \mathbb{E}_{R_\tau} (R_\tau(\hat{A}) - Q(\hat{A})) \\
&\quad = \frac{c}{4} \sum_{i = 1}^S q_i \wedge \sqrt{\frac{q_i}{n}} - \frac{1}{2} \sup_{\hat{A}} \sum_{i =1}^S 2^{-S}  \sum_{\tau} \mathbb{E}_{R_\tau} (R_{\tau}(i) - q_i) \mathbbm{1}(i\in \hat{A}) \\
& \quad =  \frac{c}{4} \sum_{i = 1}^S q_i \wedge \sqrt{\frac{q_i}{n}}  - \frac{1}{2} \sup_{\hat{A}} \sum_{i =1}^S 2^{-S-1}  \sum_{\tau} \left( \mathbb{E}_{R_\tau} (R_{\tau}(i) - q_i) \mathbbm{1}(i\in \hat{A}) +  \mathbb{E}_{R_{\tau^i}} (R_{\tau^i}(i) - q_i) \mathbbm{1}(i\in \hat{A}) \right). 
\end{align}
For each fixed $\tau$ and $i$, we have
\begin{align}
\mathbb{E}_{R_\tau} (R_{\tau}(i) - q_i) \mathbbm{1}(i\in \hat{A}) +  \mathbb{E}_{R_{\tau^i}} (R_{\tau^i}(i) - q_i) \mathbbm{1}(i\in \hat{A}) & \leq c(q_i \wedge \sqrt{\frac{q_i}{n}}) \left| \mathbb{E}_{R_\tau} \mathbbm{1}(i \in \hat{A}) - \mathbb{E}_{R_{\tau^i}} \mathbbm{1}(i\in \hat{A}) \right | \\
& \leq c(q_i \wedge \sqrt{\frac{q_i}{n}}) V(R_{\tau}(\hat{r}_1^S),R_{\tau^i}(\hat{r}_1^S)),
\end{align} 
where $V(P,Q) = \frac{1}{2}L_1(P,Q)$ is the total variation distance, with the variational characterization $V(P,Q) = \sup_{A} |P(A) - Q(A)|
$. Here $V(P_1(\hat{p}_1^S), P_2(\hat{p}_1^S) )$ denote the total variation distance between the distributions of the empirical probabilities $\{\hat{p}_i: 1\leq i\leq S\}$ when the underlying distributions are $P_1$ and $P_2$, respectively. 

Hence,
\begin{align}
\inf_{\hat{A}} \int \mathbb{E}_{R_\tau} [\ell(\hat{A}; R_\tau, Q)]\mu(dR)& \geq  \frac{c}{4} \sum_{i = 1}^S q_i \wedge \sqrt{\frac{q_i}{n}}  - \frac{1}{2} \sum_{i = 1}^S 2^{-S-1}c (q_i \wedge \sqrt{\frac{q_i}{n}}) \sum_\tau V(R_{\tau}(\hat{r}_1^S),R_{\tau^i}(\hat{r}_1^S)) \\
& \geq   \frac{c}{4} \sum_{i = 1}^S q_i \wedge \sqrt{\frac{q_i}{n}} \left( 1- \sup_{\tau,i} V(R_{\tau}(\hat{r}_1^S),R_{\tau^i}(\hat{r}_1^S))  \right) \\
\end{align}
Since
\begin{align}
\sup_{\tau,i} V(R_{\tau}(\hat{r}_1^S),R_{\tau^i}(\hat{r}_1^S)) & = V(\spo(n (q_i - c (q_i \wedge \sqrt{\frac{q_i}{n}}))),\spo(n (q_i +c( q_i \wedge \sqrt{\frac{q_i}{n}}))) ). 
\end{align}

It was shown in~\cite{adell2006exact} that for $t,x\geq 0$, 
\begin{align}
V(\spo(t),\spo(t+x)) & \leq \min\left\{ 1-e^{-x}, \sqrt{\frac{2}{e}}(\sqrt{t+x}-\sqrt{t}) \right \} \\
& \leq \min\left\{ 1-e^{-x}, \sqrt{\frac{2}{e}} \frac{x}{\sqrt{2t+x}} \right \}. 
\end{align}
Hence,
\begin{align}
 \sup_{\tau,i} V(R_{\tau}(\hat{r}_1^S),R_{\tau^i}(\hat{r}_1^S)) & \leq \frac{2c}{\sqrt{e}}. 
\end{align}

If we take $c \leq \frac{\sqrt{e}}{4}$, it is easy to see that $\frac{2c}{\sqrt{e}} \leq \frac{1}{2}$. Hence, we have under the Poisson model
\begin{align}
\inf_{\hat{A}} \int \mathbb{E}_{R_\tau} [\ell(\hat{A}; R_\tau, Q)]\mu(dR) & \geq \frac{c}{8} \sum_{i = 1}^S q_i \wedge \sqrt{\frac{q_i}{n}}
\end{align}

Denote the minimax regret under the multinomial model with sample size $n$ as $R(S,n,Q)$, it follows from Lemma~\ref{lemma.poissonmultinomial} and Lemma~\ref{lemma.bayesregret} that under the multinomial model,
\begin{align}
R(S,n(1-\epsilon)/2,Q) & \geq  \frac{c}{8} \sum_{i = 1}^S q_i \wedge \sqrt{\frac{q_i}{n}}  - \left( 1 + c \sum_{i = 1}^S q_i \wedge \sqrt{\frac{q_i}{n}} \right) \mu((\mathcal{D}_0(S,\epsilon))^c) - e^{-n(1-\epsilon)/8} - \frac{3\epsilon}{4},
\end{align}
where $\mu$ is the distribution that assigns equal probability to each vector $R_\tau$ defined in~(\ref{eqn.ptaudefine}), and
\begin{align}
\mathcal{D}_0(S,\epsilon) = \left\{ P = (p_1,p_2,\ldots,p_S): p_i\geq 0, \left| \sum_{i = 1}^S p_i -1 \right| <\epsilon \right \}.
\end{align}
It follows from Lemma~\ref{lem_hoeffding} that 
\begin{align}
 \mu((\mathcal{D}_0(S,\epsilon))^c) & \leq 2 e^{- \frac{2\epsilon^2}{ 4c^2 \cdot \sum_{i = 1}^S q_i^2 \wedge \frac{q_i}{n}}} \\
 & \leq 2e^{-\frac{n\epsilon^2}{2c^2}}.
\end{align}
Taking $\epsilon = c\sqrt{\frac{2 \ln n}{n}}$, we have
\begin{align}
R(S,n(1-\epsilon)/2,Q) & \geq  \frac{c}{8} \sum_{i = 1}^S q_i \wedge \sqrt{\frac{q_i}{n}}  - \left( 1 + c \sum_{i = 1}^S q_i \wedge \sqrt{\frac{q_i}{n}} \right) \frac{2}{n} - e^{-n(1-c\sqrt{(2\ln n)/n})/8} - \frac{3c}{4}\sqrt{\frac{2\ln n}{n}} \\
&  \gtrsim \sum_{i = 1}^S q_i \wedge \sqrt{\frac{q_i}{n}},
\end{align}
when $ \sqrt{\ln n}  \ll \sum_{i = 1}^S \sqrt{q_i} \wedge \sqrt{n} q_i$. 
\end{enumerate}

\subsection{Proof of Theorem~\ref{thm.iidcompression}}

We prove the lower bound. We have
\begin{align}
\inf_Q \sup_{P \in \mathcal{D}_0(S)} D(P_{X^n} \| Q_{X^n}) & = \inf_Q \sup_{\pi} \int D(P_{X^n} \| Q_{X^n})  \pi(dP) \\
& \geq \sup_{\pi} \inf_Q \int D(P_{X^n} \| Q_{X^n})  \pi(dP) \\
& = \sup_{\pi} \int D(P_{X^n} \| \int P_{X^n} \pi(dP) )  \pi(dP) \\
& = I(P; X^n),
\end{align}
where $\pi$ is a prior distribution on the space of memoryless sources $\mathcal{D}_0(S)$, and we have used the fact that min-max is an upper bound on max-min, and the Bayes action under the KL divergence loss is the expectation. The quantity $I(P;X^n)$ denotes the mutual information between the random distribution $P\in \mathcal{D}_0(S), P\sim \pi$ and the empirical observations $X^n$, which satisfies $X_1,X_2,\ldots,X_n|P \stackrel{\text{i.i.d.}}{\sim} P$. 

We choose the prior $\pi$ to be the uniform distribution on the simplex in $\mathbb{R}^{S}$, which is of dimension $S-1$. It follows from the data processing inequality that for the empirical distribution $\hat{P} = \hat{P}(X^n)$, we have
\begin{align}
I(P;X^n) & \geq I(P;\hat{P}) \\
& = h(P) - h(P|\hat{P}) \\
& = h(P) - h(P-\hat{P}|\hat{P}) \\
& \geq h(P) - h(P-\hat{P}). 
\end{align}
Since $\pi$ is the uniform distribution, we have $h(P) = \lg \frac{1}{(S-1)!} $. Since the empirical distribution satisfies that $\mathbb{E} \| P - \hat{P} \|^2 = \mathbb{E}_\pi [\mathbb{E} \| P - \hat{P}\|^2|P] \leq \frac{1}{n}$, it follows from the fact that the Gaussian distribution maximizes differential entropy with the same second moment that
\begin{align}
h(P-\hat{P}) & \leq \frac{S-1}{2} \lg \left( \frac{2\pi e}{n(S-1)} \right). 
\end{align}

Putting things together, using the fact that for any positive integer $n$, $n!\leq \sqrt{2\pi n} (n/e)^n e^{1/(12n)}$, 
\begin{align}
\inf_Q \sup_{P \in \mathcal{D}_0(S)} D(P_{X^n} \| Q_{X^n}) &  \geq \lg \frac{1}{(S-1)!} + \frac{S-1}{2} \lg \left( \frac{n(S-1)}{2\pi e} \right) \\
& \geq  \lg \left(  \frac{1}{\sqrt{2\pi(S-1)} e} \left( \frac{e}{S-1} \right)^{S-1} \right) +\frac{S-1}{2} \lg \left( \frac{n(S-1)}{2\pi e} \right) \\
& = -\lg \left( e\sqrt{2\pi (S-1)} \right) + \frac{S-1}{2} \lg \left( \frac{e^2}{(S-1)^2} \right) + \frac{S-1}{2} \lg \left( \frac{n(S-1)}{2\pi e} \right) \\
& = \frac{S-1}{2} \lg \left( \frac{n e}{2\pi (S-1)} \right)  -\lg \left( e\sqrt{2\pi (S-1)} \right) . 
\end{align}
Hence, if $S = \alpha n$, where $0<\alpha<\frac{e}{2\pi}$, we have
\begin{align}
\inf_Q \sup_{P \in \mathcal{D}_0(S)} \frac{1}{n} D(P_{X^n} \| Q_{X^n}) & \geq \frac{\alpha}{2} \lg \left( \frac{e}{2\pi \alpha} \right) -O \left(\frac{\lg n}{n} \right). 
\end{align}

\section{Proofs of auxiliary lemmas}\label{sec.proofsofauxiliarylemmas}

\subsection{Proof of Lemma~\ref{lemma.poissondifference}}
We prove the first statement first. Since $0\leq \lambda_2 \leq n, \lambda_2 \geq \lambda_1$, we have
\begin{align}
(\lambda_2 - \lambda_1) P(X\geq \lambda_2) & \leq (\lambda_2 - \lambda_1) P(X - \lambda_1 \geq \lambda_2 - \lambda_1)\\
& \leq (\lambda_2-\lambda_1) \frac{\mathbb{E}|X - \lambda_1|}{\lambda_2 - \lambda_1 }\\
& \leq \mathbb{E}|X - \lambda_1|. 
\end{align}
It is clear that $\mathbb{E}|X-\lambda_1|\leq \sqrt{\mathbb{E}(X - \lambda_1)^2} \leq \sqrt{\lambda_1} \leq \sqrt{\lambda_2}$. If $\lambda_1<1$, we also have
\begin{align}
\mathbb{E}|X - \lambda_1| & = \lambda_1 P(X = 0) + \sum_{j = 1}^\infty P(X = j)(j-\lambda_1) \\
& = \lambda_1 P(X = 0) + \mathbb{E}[X] - \lambda_1 (1-P(X = 0))\\
& = \lambda_1\left( P(X = 0) + 1 - 1 + P(X = 0) \right)\\
& = 2\lambda_1 e^{-\lambda_1} \\
& \leq 2\lambda_2. 
\end{align}

The first statement is proved. Now we consider the second statement. We used the classical splitting operation~\cite{Jiao--Han--Weissman2016l1distance} to represent random variable $X$ as 
\begin{align}
X = Y+Z,
\end{align}
where $Y \sim \spo(\lambda_2), Z \sim \spo(\lambda_1 - \lambda_2)$ if $X \sim \spo(\lambda_1)$, and $Y \sim \mathsf{B}(n, \frac{\lambda_2}{n}), Z \sim \mathsf{B}(n, \frac{\lambda_1 - \lambda_2}{n})$ if $X \sim \mathsf{B}(n, \frac{\lambda_1}{n})$. Note that in the Poisson case, we have $Y$ is independent of $Z$, and in the binomial case, we no longer have independence, but the random variables $Y,Z$ are negatively associated~\cite[
Cor. 8]{Jiao--Venkat--Weissman2014MLE}. 

Then, 
\begin{align}
(\lambda_1 - \lambda_2) P(X \leq \lambda_2) & = (\lambda_1 - \lambda_2) \sum_{j = 0}^\infty P(Z = j, Y\leq \lambda_2 - j) \\
& = (\lambda_1 - \lambda_2) \mathbb{E} \left[  \sum_{j = 0}^\infty \mathbbm{1}(Z = j, \lambda_2 - Y \geq  j) \right] \\
& \leq  (\lambda_1 - \lambda_2) \mathbb{E} \left[ \mathbbm{1}(Z = 0) +  \sum_{j = 1}^\infty 
\mathbbm{1}(Z = j) \frac{(\lambda_2-Y)_+}{j} \right] \\
& =  (\lambda_1 - \lambda_2) \mathbb{E} \left[ \mathbbm{1}(Z = 0) +   \sum_{j = 1}^\infty \mathbbm{1}(Z = j) \frac{(\lambda_2-Y)_+}{Z} \right] \\
&  \leq  (\lambda_1 - \lambda_2) P(Z = 0) + (\lambda_1 - \lambda_2) \mathbb{E} \left[ \frac{(\lambda_2-Y)_+}{Z \vee 1} \right]
\end{align}
Using the negative association property of $Y$ and $Z$, we have
\begin{align}
(\lambda_1 - \lambda_2) P(X \leq \lambda_2) & \leq  (\lambda_1 - \lambda_2) P(Z = 0) + (\lambda_1 - \lambda_2) \mathbb{E}[(\lambda_2 - Y)_+] \mathbb{E} \left[ \frac{1}{Z \vee 1} \right] \\
& \leq (\lambda_1 - \lambda_2) e^{-(\lambda_1 - \lambda_2)} + \mathbb{E}[(\lambda_2 - Y)_+] \mathbb{E}\left[ \frac{(\lambda_1 - \lambda_2)}{Z \vee 1}\right] \\
& \leq \frac{1}{e} + C  \mathbb{E}[(\lambda_2 - Y)_+] \\
& \leq \frac{1}{e} + C \mathbb{E}|\lambda_2 - Y| \\
& \leq \frac{1}{e} + C \sqrt{\mathbb{E}(\lambda_2 - Y)^2} \\
& \leq \frac{1}{e} + C \sqrt{\lambda_2} \\
& \leq (C + \frac{1}{e}) \sqrt{\lambda_2},
\end{align}
where $C>0$ is the universal constant in Lemma~\ref{lemma.poissoninversebound}, and we used the assumption that $\lambda_2\geq 1$.

\subsection{Proof of Lemma~\ref{lemma.poissonmultinomial}}

For an arbitrary $\delta>0$, it follows from the definition of the minimax regret that there exists a near-minimax decision regime $\hat{A}$ for every sample size $n$ such that
\begin{align}
\sup_{R\in \mathcal{D}_0(S,0)}\mathbb{E}_R [\ell(\hat{A};R,Q)] & \leq R(S,n,Q) + \delta. 
\end{align}

We now use this decision rule under the Poisson model. For any distribution $R\in \mathcal{D}_0(S,\epsilon)$ under the Poisson model, let $n\cdot \hat{r}_i \sim \spo(nr_i)$, $n' = \sum_{i = 1}^S n \hat{r}_i \sim \spo(n \sum_{i = 1}^S r_i)$ and $B = \sum_{i = 1}^S r_i$. It follows from the fact that $R \in \mathcal{D}_0(S,\epsilon)$ that $|B-1|\leq \epsilon$.  We have under the Poisson model
\begin{align}
\mathbb{E}_R [\ell(\hat{A};R,Q)] & = \mathbb{E}\left[\frac{1}{2} \left(  Q(\hat{A}) - R(\hat{A}) + \frac{1}{2} L_1(R,Q) \right)    \right] \\
& = \frac{1}{2} \mathbb{E} \left[  \sum_{i = 1}^S (q_i - r_i) \mathbbm{1}(i\in \hat{A}) \right] + \frac{1}{4}L_1(R,Q) \\
& = \frac{1}{2} \mathbb{E} \left[ \sum_{i = 1}^S (q_i - r_i/B + r_i/B - r_i) \mathbbm{1}(i \in \hat{A}) \right] + \frac{1}{4}L_1\left( \frac{R}{B}, Q \right) - \frac{1}{4}L_1\left( \frac{R}{B},Q \right) + \frac{1}{4}L_1(R,Q) \\
& \leq \frac{1}{2} \mathbb{E} \left[  \sum_{i = 1}^S (q_i - r_i/B) \mathbbm{1}(i\in \hat{A}) \right]  + \frac{1}{4}L_1 \left( \frac{R}{B},Q \right) + \frac{1}{2}\mathbb{E} \left[ \sum_{i = 1}^S (r_i/B - r_i) \mathbbm{1}(i\in \hat{A}) \right] + \frac{1}{4}\left| L_1(R,Q) - L_1\left( \frac{R}{B},Q \right) \right| \\
& \leq \frac{1}{2} \mathbb{E} \left[  \sum_{i = 1}^S (q_i - r_i/B) \mathbbm{1}(i\in \hat{A}) \right]  + \frac{1}{4}L_1 \left( \frac{R}{B},Q \right) + \frac{\epsilon}{2} + \frac{\epsilon}{4} \\
& =  \sum_{m = 0}^\infty \left( \mathbb{E}  \left[\ell(\hat{A};R,Q) \Bigg | n' =m \right] P(n' = m)  \right) + \frac{3\epsilon}{4} \\
& \leq  \left( \sum_{m = 0}^\infty R(S,m,Q) P(n' = m) \right) +\delta + \frac{3\epsilon}{4} \\
& \leq    1 \cdot P(n'\leq n(1-\epsilon)/2) + R(S, n(1-\epsilon)/2, Q) \cdot P(n' \geq n(1-\epsilon)/2)  + \delta + \frac{3\epsilon}{4} \\
& \leq R(S,n(1-\epsilon)/2, Q) +  P( \spo(n(1-\epsilon)) \leq n(1-\epsilon)/2) + \delta+ \frac{3\epsilon}{4} \\
& \leq R(S,n(1-\epsilon)/2, Q) +  e^{-n(1-\epsilon)/8} + \delta + \frac{3\epsilon}{4},
\end{align}
where we used the fact that conditioned on $n' = m$, the random vector $(n\hat{r}_1, n\hat{r}_2,\ldots,n\hat{r}_S)$ follows the multinomial distribution with parameter $(n, \frac{R}{\sum_{i = 1}^S r_i})$, the monotonicity of $R(S,m,Q)$ as a function of $m$, the fact that $R(S,m,Q)\leq 1$, and Lemma~\ref{lemma.poissontail}. 

Taking supremum over the distribution $R\in \mathcal{D}_0(S,\epsilon)$ and using the arbitrariness of $\delta$, we obtain
\begin{align}
R(S,n(1-\epsilon)/2, Q) & \geq R_P(S,n,Q,\epsilon) - e^{-n(1-\epsilon)/8} - \frac{3\epsilon}{4}. 
\end{align}

\subsection{Proof of Lemma~\ref{lemma.bayesregret}}

Define the conditional prior $\pi$ by 
\begin{align}
\pi(A) = \frac{\mu(A \cap \mathcal{D}_0(S,\epsilon))}{\mu(\mathcal{D}_0(S,\epsilon))},
\end{align}
we consider the Bayes decision regime $\hat{A}_\pi$ under prior $\pi$ and the corresponding Bayes regret $R_B(S,n, Q, \pi)$. Since $R_B(S,n,Q,\mu)$ is the Bayes regret under $\mu$, applying $\hat{A}_\pi$ will result in at least as much as regret:
\begin{align}
R_B(S,n,Q,\mu) & \leq \int   \mathbb{E}[\ell(\hat{A}_\pi;R,Q)] \mu(dR) \\
& \leq \int_{\mathcal{D}_0(S,\epsilon)} \mathbb{E}[\ell(\hat{A}_\pi;R,Q)] \mu(dR) + \int_{(\mathcal{D}_0(S,\epsilon))^c} \mathbb{E}[\ell(\hat{A}_\pi;R,Q)] \mu(dR) \\
& \leq \mu(\mathcal{D}_0(S,\epsilon)) \int_{\mathcal{D}_0(S,\epsilon)} \mathbb{E}[\ell(\hat{A}_\pi;R,Q)] \pi(dR) + C \mu((\mathcal{D}_0(S,\epsilon))^c) \\
& \leq R_B(S,n,Q,\pi) + C \mu((\mathcal{D}_0(S,\epsilon))^c) \\
& \leq R_P(S,n,Q,\epsilon) + C \mu((\mathcal{D}_0(S,\epsilon))^c),
\end{align}
where we use the fact that the Bayes regret is a lower bound of the minimax regret.

\bibliographystyle{IEEEtran}
\bibliography{di}

\end{document}